\documentclass[preprint2]{aastex}

\slugcomment{Not to appear in Nonlearned J., 45.}

\shorttitle{Gamma-ray and optical oscillations in PKS 0537-441}
\shortauthors{A. Sandrinelli, S. Covino, A. Treves}

\begin{document}


\title{Gamma-ray and optical oscillations in the blazar PKS 0537-441}


\author{A. Sandrinelli}
\affil{\textit{Universit\`a degli Studi del$\l'$Insubria, Via Valleggio 11, I-22100 Como, Italy\\
INAF - Osservatorio Astronomico di Brera, Via Emilio Bianchi 46, I-23807 Merate, Italy.
}}
\email{asandrinelli@yahoo.it}

\author{S. Covino}
\affil{\textit{
INAF - Osservatorio Astronomico di Brera, Via Emilio Bianchi 46, I-23807 Merate, Italy}}

\author{A. Treves}
\affil{\textit{Universit\`a degli Studi del$\l'$Insubria, Via Valleggio 11, I-22100 Como, Italy\\
INAF - Osservatorio Astronomico di Brera, Via Emilio Bianchi 46, I-23807 Merate, Italy\\
INFN - Istituto Nazionale di Fisica Nucleare, Sezione Trieste - Udine, \\
 Via Valerio 2, I-34127 Trieste, Italy 
 }}

\altaffiltext{1}{......}
\altaffiltext{2}{.......}

\begin{abstract}

We have considered the \textit{Fermi} $\gamma$-ray light curve of the blazar PKS 0537-441 
during a high state extending from 2008/08/10 to  2011/08/27. 
The periodogram exhibits a peak at  T $\sim$ 280 d, with a significance of  $\sim$ 99.7 \%.  
A peak of similar relevance at $\frac{1}{2}$ T is found in the optical light curves. 
 Considering the entire duration of the \textit{ Fermi}  light curve 2008-2015, no significant 
 peak is revealed, while the optical one remains meaningful.  
 Comparing with recent observations of PKS 2155-304 and PG 1553+113 it seems that 
 month-year oscillations  can characterize some blazars. 
 Month-scale oscillations can also show up only during phases of enhanced or bursting
  emission like in the case of PKS 0537-441.

\end{abstract}


\section{Introduction}

 \begin{figure*}[t!]
\centering
\includegraphics[trim=0.8cm 1cm 0cm 1cm,clip,width=1.4\columnwidth]{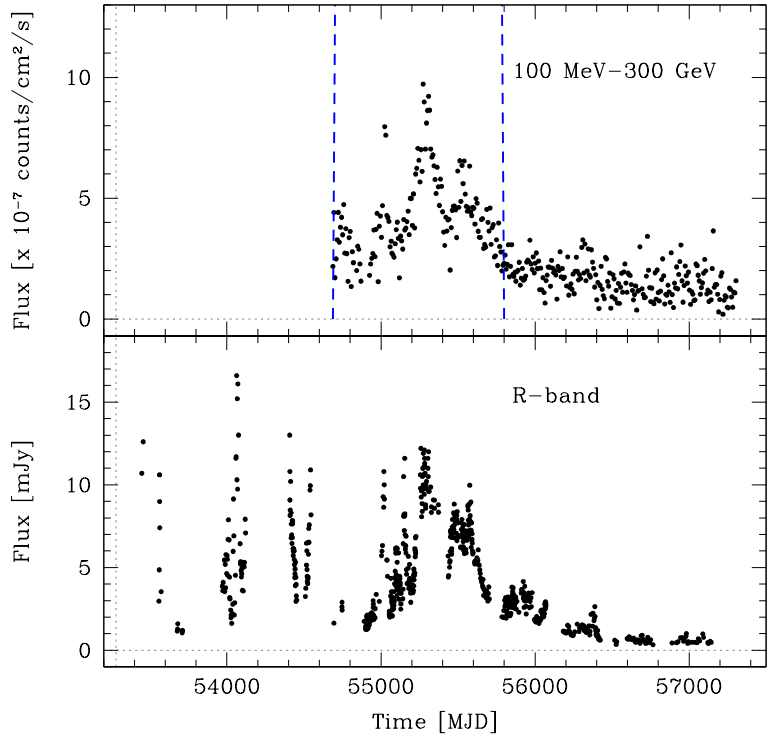}  
              \caption{\label{lc0}  
                  Weekly averaged {Fermi} $\gamma$-ray  light curve of PKS 0537-441
                  in the 100 MeV - 300 GeV  energy range (\textit{top panel}). 
	Nightly averaged REM and SMARTS  light curve in R band
	is  also reported (\textit{bottom panel}). Vertical lines identify the $\gamma$-ray \textit{high} state of the source.
	Errors are omitted for readability.
             }
\end{figure*}

Blazars are jet dominated active galactic nuclei   with the jet pointing in the observer direction.  
A characteristic of the class is a large variability on various timescales and in all spectral bands 
\citep[see for a recent review][]{Falomo2014}.  
Periodicities have been searched for, since the discovery of the first members of the class.  
No convincing period was found, with the possible exception of the case of OJ 287, 
where the claim of a 12 year period \citep[e.g.][]{Sillanpaa1988} appears rather robust,
 but not uncontroversial \citep[e.g.][]{Hudec2013} .

\begin{figure*}[t!]
\centering\
\includegraphics[trim=0.8cm 1cm 0cm 1cm,clip,width=1.4\columnwidth]{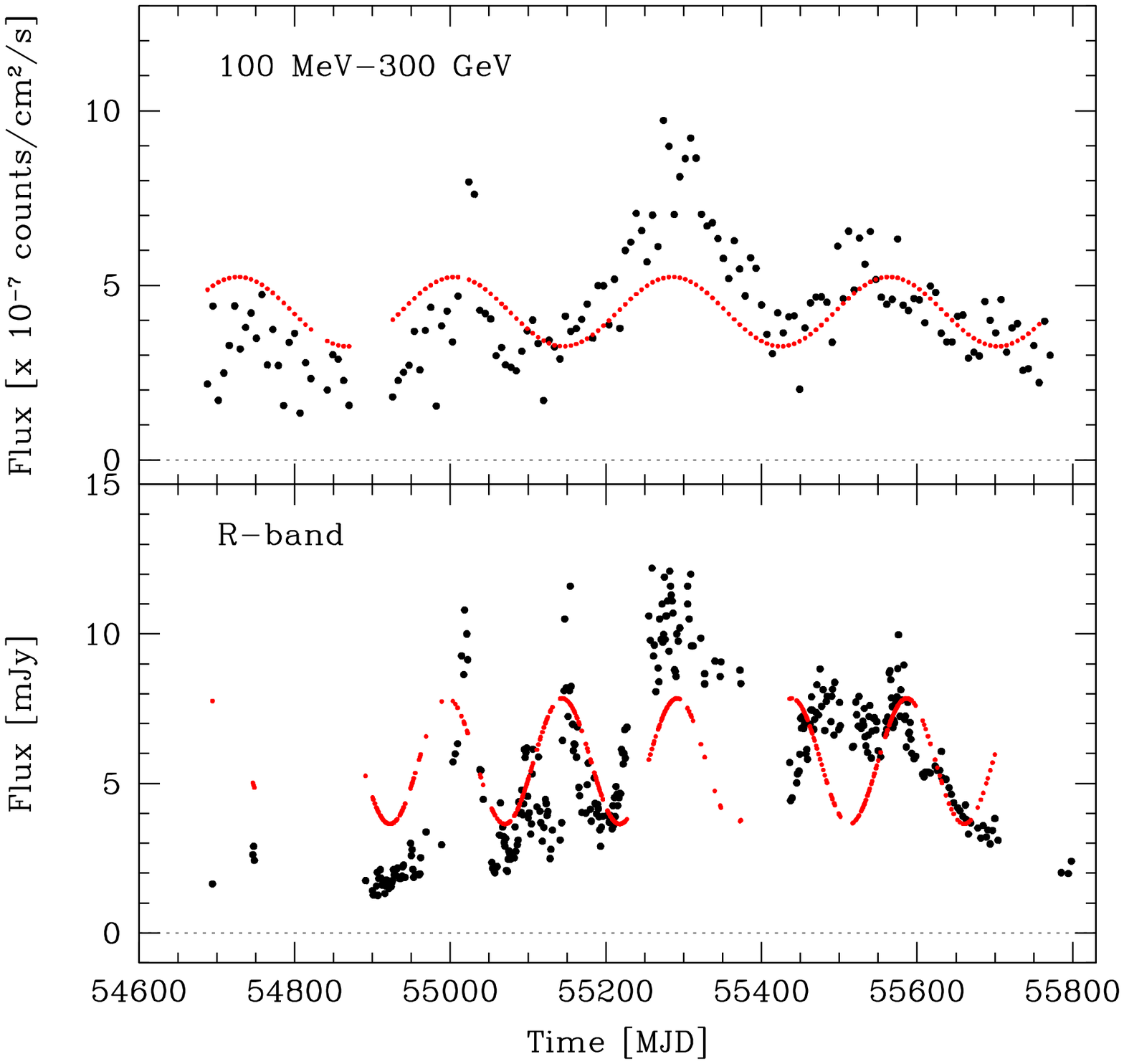}  
              \caption{\label{lc}  
   Enlargement of the high state of the source, see Figure \ref{lc0}.
   The red lines are the sinusoidal artificial models referring to the $\gamma$-ray period
   $T_0$ = 280 d and  to the optical period $T_1$ = 148 d.
	Errors are omitted for readability.
             }
\end{figure*}

The situation has changed significantly in recent years, because it has become clear
that blazars are the main constituents of the extragalactic $\gamma$-ray sky and the \textit{Fermi} mission, 
since its launch has monitored the entire celestial sphere every 3 hours.
The consequence is that $\gamma$-ray light curves of blazars are easily available.  
The   \textit{Fermi} Large Area Telescope (LAT) observations and light curve extraction procedures 
are described in detail e.g. in  \cite{Abdo2010}.
The \textit{Fermi} Collaboration 
provides  daily and weekly flux light curves of monitored sources 
of interest in  automated analysis\footnote{http://fermi.gsfc.nasa.gov/ssc/data/access/lat/msl\_l/}.
 The available energy ranges are 1-300 GeV, 300 MeV-1GeV, 100 MeV-300 GeV.
At the same time robotic optical telescopes have become rather common, 
so that the monitoring time dedicated to blazars has  substantially increased.

\cite{Sandrinelli2014a} examined the  \textit{Fermi}  light curves of the prototypical 
BL Lac object PKS 2155-304  (redshift z=0.116,  R magnitude $\sim$13) 
and discovered a significant periodicity
 of T $\sim$ 630 d, which is twice that obtained by \cite{Zhang2014} 
 making a collection of all the published optical photometry of the source in 35 years.  
 The existence of the optical periodicity  was confirmed by our independent 
 Rapid Eye Mounting Telescope  \citep[REM,][]{Zerbi2004, Covino2004} photometry,
 which covered the source from  2008 to 2015/03 \citep{Sandrinelli2014b,Sandrinelli2015}.
 Recent papers \citep{Ackermann2015,Hughes2015}  considered the \textit{Fermi}  
light curves of another bright  BL Lac object  PG 1553+113  (z$\sim$0.4,  R$\sim$14).
A periodicity of 2.18 years is found with 
interesting significance $S\gtrsim$ 99\%, which shows up clearly  also
in the R-monitoring.  In \cite{Sandrinelli2015}  we examined the $\gamma$-ray and optical
light curves of  PKS 0537-441, OJ 287, 3C 279,  PKS 1510-089,  PKS 2005-489 and  
PKS 2155-304.  While the power spectra indicated several peaks  in all sources,
 sometimes at related frequencies in $\gamma$-ray and optical filters, 
 with the exception PKS 2155-304  the significance is not very high. 
Short duration oscillatory patterns have been reported  in various light curves of blazars,
 both in the optical and X-rays, e.g.
in PKS 2155-304 \citep{Urry1993,Lachowicz2009}, 
OJ 287 \citep{Kinzel1988},
S5 0716+714 \citep{Rani2010}. 
Year time scale modulations  were found in 
GB6  J1058+5628 \citep{Nesci2010},  
 and also in radio bands, e.g.
  J1359+4011 \citep{King2013}, 
 and PKS 1156+295 \citep{Wang2014} 
Quasi-periodical   outbursts were observed occurring in
e.g. S5 0716+714 \cite{Raiteri2003}	 
and AO 0235+164 \citep [][and references therein]{Raiteri2008}.
 
 In this paper we refocus the case of PKS  0537-441, concentrating  only on a rather high state.

\begin{figure}[h!]
\centering
\includegraphics[trim=0.8cm 1cm 0cm 0.2cm,clip,width=1\columnwidth]{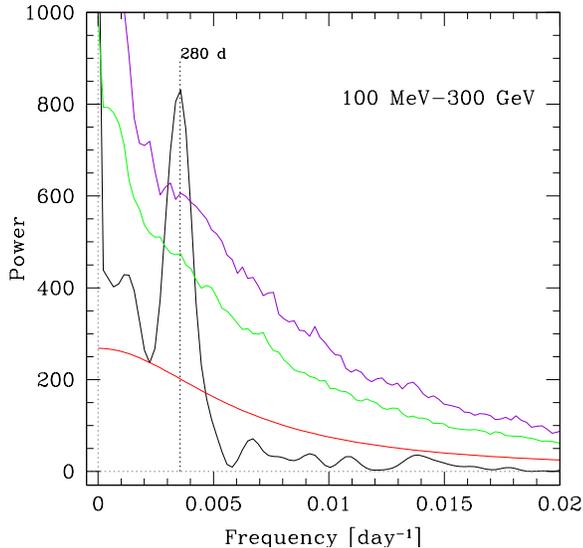}  
              \caption{\label{redfitg1_55570_r62}  
		Periodogram  (black line) of the active phase of  PKS 0537-441  
		(from 2008/08/10 to   2011/08/27) obtained from  the  \textit{Fermi} 100 MeV-300 GeV data.
		 The curves  from the bottom are:  red-noise spectrum (red curve),
		   95\%   (2$\sigma$, green curve) and 99\% (2.5$\sigma$, purple curve) Monte Carlo 
		 significance levels after 1000 simulations. 
               	The period in days corresponding  to the prominent peak is marked.
		}
\end{figure}

\section{PKS 0537-441}

It is a very well studied object (z=0.896,  R$\sim$14), with characteristics intermediate 
between Optically Violently Variable Quasars and BL Lacs. 
It was observed in all spectral bands from radio to GeV $\gamma$-rays.  
Detailed information on the source can be  found in \cite{Pian2007} and \cite{D'Ammando2013}. 
In Figure \ref{lc0} we report the \textit{Fermi} $\gamma$-ray light curve with one-week 
integration 
in the 100 MeV-300 GeV energy band, as provided by the \textit{Fermi} LAT team (see above).
We report also the R-band light curve obtained combining  REM and Small \& Moderate 
Aperture Research Telescope System data 
 \citep[SMARTS\footnote{\texttt{http://www.astro.yale.edu/smarts/glast/home.php}},][]{Bonning2012}. 
 Details can be found in \cite{Sandrinelli2015}.

\section{Search for periodicities}

Our procedure for searching for regular oscillations
follows closely that of \cite{Sandrinelli2015}
 and is based on the scheme proposed by \cite {Schulz2002}.  
 Shortly, 1) it yields  the Lomb and Scargle  \citep{Scargle1982} periodograms
 also accounting for unevenly  spaced photometry, when it is the case, 2) it models  the red noise
 with a first-order autoregressive process as  null hypothesis of stochastic events,  
 3) it provides the bias-corrected periodograms and significance levels  using Monte Carlo simulations.  
 
 The analysis of the entire $\gamma$-ray light curve  from 54688 MJD (2008/08/10)  
 to  57304 MJD  (2015/10/09)  
 gives a peak of low significance  at  351 d ($S\sim$ 90\%).
 Interestingly enough in the  R, J, K curves there 
 are peaks of  significance  $\gtrsim$ 99\% at $\sim$150 d
 \citep[see also Figure 8 in][for both $\gamma$-ray  and optical  power spectra]{Sandrinelli2015}. 
 The absence of significant $\gamma$-ray peaks is consistent with the results of 
 \cite{D'Ammando2013},  who considered the  \textit{Fermi} light curve only up to  
55290 MJD (2010/04/04),  and to the standard 
time analysis\footnote{\texttt{http://fermi.gsfc.nasa.gov/ssc/data/access/lat/\\
4yr\_catalog/ap\_lcs.php}} 
provided by the  \textit{Fermi} collaboration for the 2008-2015 light curve.

An examination of the $\gamma$-ray light curve (100 MeV-300 GeV) reported in Figure \ref{lc0} 
indicates a rather high $\gamma$-ray state of the source  from $\sim$ 54688 MJD (2008/08/10)
to $\sim$ 55800 MJD (2011/08/27),
 followed by a lower state.
The same is found in the 300 MeV-1 GeV and 1 GeV-300 GeV $\gamma$-ray 
light curves. 
We note that the state of the source remains low untill the last available observations
of \textit{Fermi}.
Moreover the high state is only partially contained in \cite{D'Ammando2013} $\gamma$-ray light curve.
Both high and low $\gamma$-ray states have  clear counterparts in the optical bands
 \citep[see also][]{Sandrinelli2015}.  
 The historical optical light curves, in fact,  show that high states of duration of years
are not infrequent \citep{Pian2007}.  
This kind of pattern is present also in the blazar 3C 454.3 \citep[][and references therein]{Abdo2011}.
Here we concentrate on the high $\gamma$-ray state of PKS 0537-441.
(Figure \ref{lc}).  
The Lomb-Scargle analysis, as specified above, shows a peak at 279 d with 99\% significance. 
We also proceeded 
in splitting  the light curve in  
2 segments, which are overlapping  for the 50\%
of their length, adopting the Welch-overlapped-segment-averaging \citep[WOSA,][]{Welch1967}
to 
enhance enduring quasi-periodicities.
The obtained  periodogram is given in Figure \ref{redfitg1_55570_r62},
where the  peak at $T_0$ =280 $\pm$ 39 d with $S\sim 3\sigma$ (99.7 \%) is apparent. 
The errors  are evaluated  following  \cite{Schwar1991},
who applies the Mean Noise Power Level (MNPL) method.
The 1$\sigma$ confidence interval on the investigated period T 
is the width of the peak  at the $p$-$n$  level, 
where $p$ is the height of the peak and $n$ is the mean noise power level
in the vicinity of T.  We adapted  the approach to our data
considering the mean local 1st-order autoregressive red-noise power level \citep{Schulz2002}.

\begin{figure}[h!]
\centering
\includegraphics[trim=0.8cm 1cm 0cm 0.2cm,clip,width=1\columnwidth]{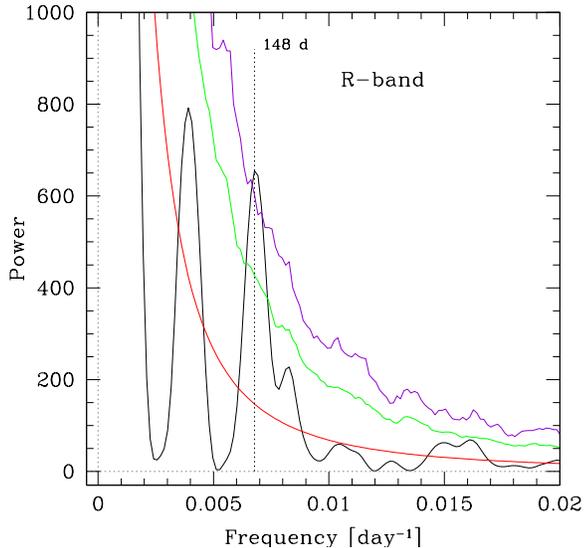}  
              \caption{\label{redfitR_54688_55800}   
		Periodogram  (black line) of the active phase of  PKS 0537-441  
		(from 2008/08/10 to   2011/08/27) obtained from  the REM+SMARTS R-band photometry.
		 The curves  from the bottom are:  red-noise spectrum (red curve),
		   95\%   (2$\sigma$, green curve) and 99\% (2.5$\sigma$, purple curve) Monte Carlo 
		 significance levels after 1000 simulations. 
               	The period in days corresponding  to the prominent peak is marked.
		}
\end{figure}

\begin{figure*}
\centering
\includegraphics[width=.4\textwidth]{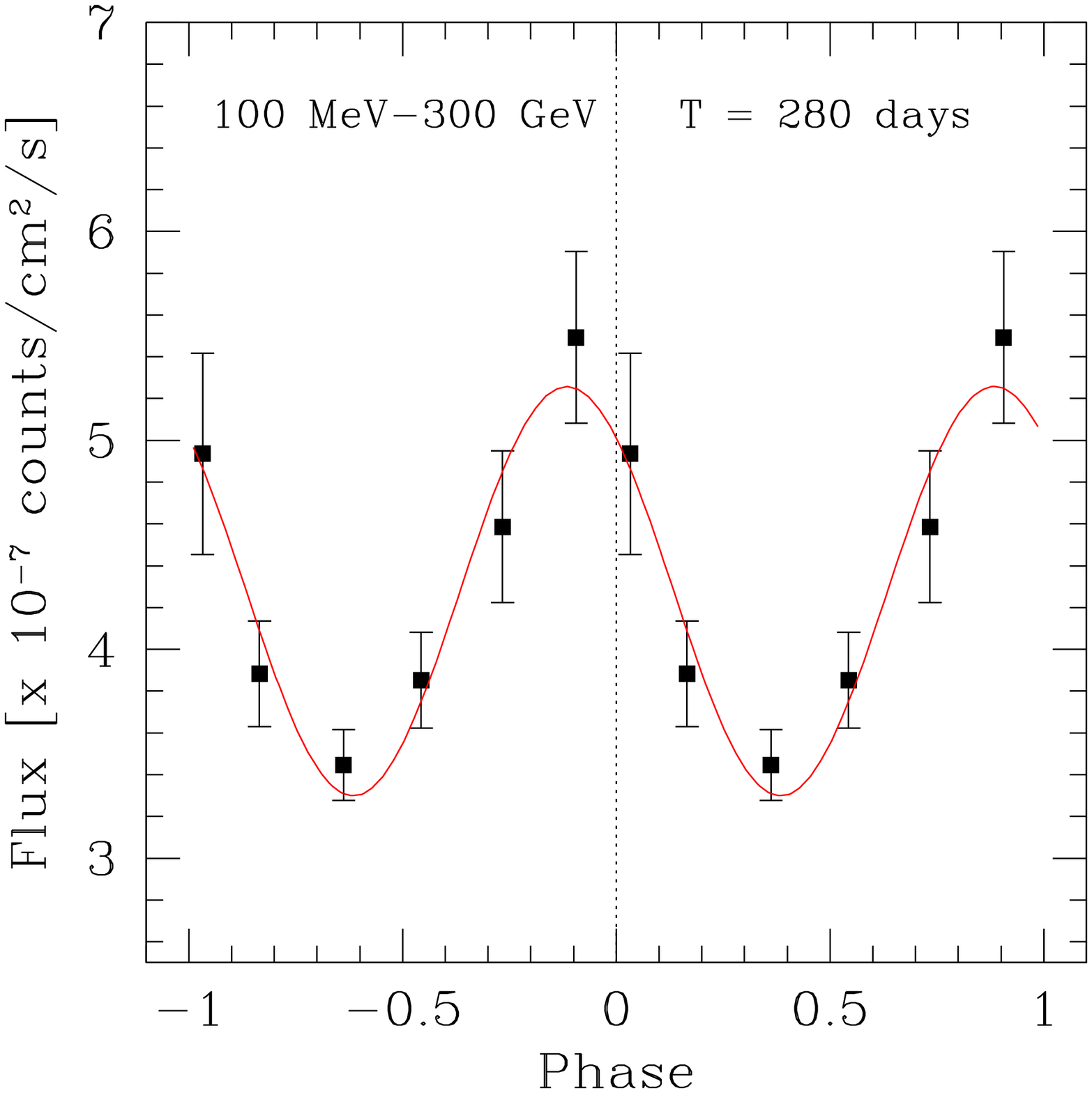}  
\includegraphics[width=.4\textwidth]{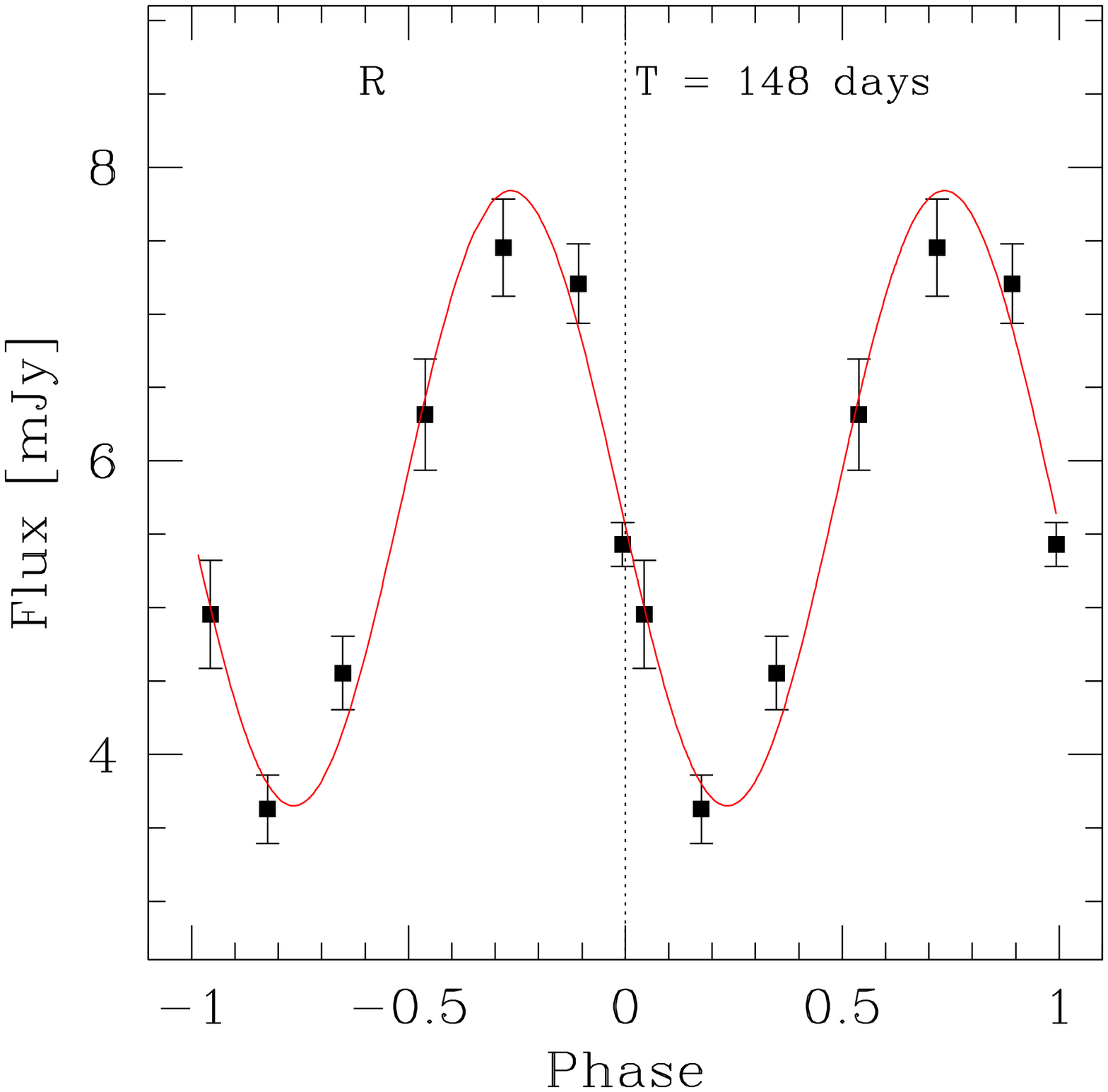}  
              \caption{\label{fold}   
		Folded light curves  for the active phase of PKS 0537-441. 
		Vertical bars are the standard errors of the average. }
\end{figure*}

The periodogram referring to the R light curve  is reported in Figure \ref{redfitR_54688_55800}.
It is important to note that the peak at $T_1$= 148 $\pm$ 17 d has a $S \sim 99 \%$ and it is within the errors  
at  one half of the $\gamma$-ray period $T_0$. 
It is also compatible with the peak in the periodogram 
of the 2008-2014 observations (see introduction).
The analysis applied to the other optical bands  yields comparable results.
 In Figure \ref{lc} we report ed sinusoidal artificial light curves with  periods specified  
 above  calculated using the Vstar package\footnote{\texttt{http://www.aavso.org/vstar-overview}}. 
 The amplitude of the components are 
 A=1.0  $\cdot$ 10$^{-7}$  
 photons $\cdot$ s$^{-1}$ $\cdot$ cm$^{2}$ \ in 100 MeV-300 GeV ($T_0$=280 d), 	
and  A=2.1 mJy  in R  ($T_1$=148 d).			
Clearly the sinusoidal pattern is superposed to a strong chaotic variability.
Folded light curves for the $\gamma$-ray and optical data 
obtained  by the same package are reported in Figure \ref{fold}.

\begin{figure}[h!]
\centering
\includegraphics[width=1\columnwidth]{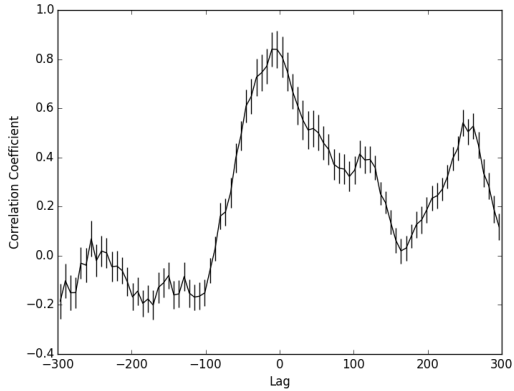}  
              \caption{\label{cc}   
		Discrete cross-correlation function (7 days bin)  between  100 MeV-300 GeV 	
		and  R-band data.				
		}
\end{figure}

The $\gamma$-ray / optical discrete cross-correlation was obtained following \cite{Edelson1988},
and it is shown in Figure \ref{cc}.
The main peak is at zero delay with full width half intensity estimated from
the left side  of $\sim$ 70 d. The (-100 d; +100 d) section is similar to that
obtained by  \cite{D'Ammando2013}. There are also two peaks  at $\sim$125 d and $\sim$ 265 d
 which are possibly related to the mentioned oscillations.

\section{Summary and Discussion}

We have concentrated on a high state of PKS 0537-441 of a duration of $\sim$ 1000 d, 
and have shown that the periodogram of the $\gamma$-ray light curve exhibits a peak
 at  $T_0\sim$ 280 d with a significance of 99.7\%.  
 At $T_1 \sim \frac{1}{2} \cdot T_0$ there is also a peak in the optical light curve with 99\% significance. 
 Because our light curve covers only $\sim$ 3 periods,  the indication is at most for a
short-lived quasi-periodicity rather than a ÒtrueÓ period.  Interestingly enough a $\gamma$-ray 
 period at twice the optical one was found also in PKS 2155-304 \citep{Sandrinelli2014a}.  
We note that no explanation for the factor 2 difference in the time scale of the optical 
 and $\gamma$-ray oscillations have been suggested thus far.
 Considering also the quasi-periodicity in PG 1553-113 (see introduction),  
 it seems to us that even if in each source the significance is not extreme, the indication 
 of  month-year oscillations in  some blazars should be considered in detail.  
 For possible preliminary interpretative  scenarios we refer to 
 \cite{Sandrinelli2014a,Sandrinelli2015} and \cite{Ackermann2015},
 where oscillations are interpreted as periodic phenomenons,
 related to a binary system of two supermassive black holes (SMBH)
 or the precession of jet-accretion disk system.  
For the case of  the observed short-lived periodicities in PKS 0537-411, 
which exhibits  regular patterns  during the $\gamma$-ray/optical 
 simultaneous flares, outbursts could be ascribed to periodic  tidal-induced disk 
  instabilities
and relaxation processes,  due, e.g., to a secondary orbiting black hole
\citep[e.g.,][]{Lehto1996,Valtonen2009} or to a binary  SMBH torque-warped
 circumbinary disk \citep[e.g.,][]{Graham2015}. 
 Disk-connected jet flow, where optical and $\gamma$-ray emissions are generally considered co-spatial \citep[e.g.,][]{Ghisellini2009},
could mimic these regular oscillations. 
 These perturbations, or different disk instabilities, could also excite  dynamical oscillatory responses  
into the accretion disk, where the presence of a number of restoring forces can be   
invoked, similarly to the 
quasi-periodic oscillations in micro quasars 
\citep[see, e.g.,][and refences therein]{Abramowicz2013}.

 If the oscillations are not related to a real periodicity, they could be due to some global 
 instabilities in the jet  
connected to its helical pattern or shocks in jets.
 Several possible explanations are discussed in 
 \cite{Marscher2014}, \cite{Godfrey2012}, \cite{Larionov2013}, \cite{Camenzind1992}
 and \cite{Marscher1985}.

  A consequence of our results is  that the \textit{Fermi} light curves of blazars should
  be carefully reconsidered.  It is possible that quasi-periodicities   do not always appear in
   the long-term  $\gamma$-ray light  curve, but  rather they show up episodically, possibly 
   in correspondence of the higher states.

\acknowledgments

We thank the referee for the  constructive comments
which enabled us to improve our paper.
This paper  made use of up-to-date SMARTS optical/near-infrared 
 light curves that are available on line.

\end{document}